\documentclass[
    twocolumn,
	prd,
	amssymb,
	preprintnumbers,superscriptaddress,
	nofootinbib]{revtex4-1}
	  
\pdfoutput=1

\usepackage{graphicx}
\usepackage{enumitem}
\usepackage{latexsym}
\usepackage{amsfonts}
\usepackage{amssymb}
\usepackage{xcolor}
\usepackage{amsmath}
\usepackage{slashed}
\usepackage{verbatim}
\usepackage{float}
\usepackage{multirow}
\usepackage{xspace}
\usepackage[normalem]{ulem}
\usepackage{hyperref}

\newcommand{\beq}{\begin{equation}}
\newcommand{\eeq}{\end{equation}}
\newcommand{\bea}{\begin{eqnarray}}
\newcommand{\eea}{\end{eqnarray}}

\allowdisplaybreaks

\begin{document}
\title{Detectable Gravitational Wave Signals from Affleck-Dine Baryogenesis}

\author{Graham White}
\email{graham.white@ipmu.jp}

\affiliation{Kavli IPMU (WPI), UTIAS, The University of Tokyo, Kashiwa, Chiba 277-8583, Japan}
\email{Author order determined by coinflip}
\author{Lauren Pearce}
\email{lpearce@psu.edu}
\affiliation{Pennsylvania State University-New Kensington, New Kensington, PA 15068}

\author{Daniel Vagie}
\email{daniel.d.vagie-1@ou.edu}
\affiliation{ Department of Physics and Astronomy, University of Oklahoma, Norman, OK 73019, USA}

\author{Alexander Kusenko}
\email{kusenko@g.ucla.edu}
\affiliation{Department of Physics and Astronomy, UCLA, Los Angeles, CA 90095, USA}
\affiliation{Kavli IPMU (WPI), UTIAS, The University of Tokyo, Kashiwa, Chiba 277-8583, Japan}

\preprint{}

\begin{abstract}
In Affleck-Dine baryogenesis, the observed baryon asymmetry of the Universe is generated through the evolution of the vacuum expectation value (VEV) of a scalar condensate.  This scalar condensate generically fragments into non-topological solitons (Q-balls).  If they are sufficiently long-lived, they lead to an early matter domination epoch, which enhances the primordial gravitational wave signal for modes that enter the horizon during this epoch.  The sudden decay of the Q-balls results in a rapid transition from matter to radiation domination, producing a sharp peak in the gravitational wave power spectrum. Avoiding the gravitino over-abundance problem favours scenarios where the peak frequency of the resonance is within the range of the Einstein Telescope and/or DECIGO.  This observable signal provides a mechanism to test Affleck-Dine baryogenesis.

\end{abstract}

\maketitle



The asymmetry between matter and anti-matter is a cornerstone puzzle of modern particle cosmology, as the Standard Model fails to provide an explanation \cite{Sakharov:1967dj,Shaposhnikov:1987tw,Dine:2003ax}. An elegant paradigm for explaining the asymmetry is the Affleck-Dine mechanism \cite{Affleck:1984fy,Dine:1995kz,Dine:2003ax,Allahverdi:2012ju}. Supersymmetric theories generically have flat directions \cite{Gherghetta:1995dv,Dine:1995kz}, which have non-zero baryon or lepton number.  During inflation, a scalar condensate generically develops in these directions, whose non-zero vacuum expectation value (VEV) spontaneously breaks C and CP.  At the end of inflation, a baryon and/or lepton asymmetry is generated as the VEV coherently evolves and the condensate fragments~\cite{Kusenko:1997si}.  These resulting clumps may be long-lived non-topological solitons (Q-balls)~\cite{Rosen:1968mfz,Friedberg:1976me,Coleman:1985ki,Kusenko:1997ad}, carrying either lepton or baryon number \cite{Kusenko:1997zq}.  This global charge is transferred to Standard Model particles when the Q-balls decay. 

However, the Affleck-Dine mechanism is generically a high-scale phenomenon which does not require a low SUSY-breaking scale, making it difficult to confirm observationally.  In this letter, we argue that a broad class of Affleck-Dine models significantly enhance the primordial gravitational wave power spectrum.  This provides a novel mechanism to test or constrain these models.

Generically, Q-balls produced through the fragmentation of the Affleck-Dine condensate are large and long-lived.  Consequently, they evolve as non-relativistic matter, and may eventually come to dominate the energy density of the Universe.  If the Q-balls decay rapidly, there is a sudden change in the equation of state for the Universe.  This results in rapidly oscillating scalar perturbations, which enhances the primordial gravitational wave spectrum from inflation.  This is analogous to the poltergeist mechanism, in which the sudden decay of black holes also enhances the gravitational wave spectrum~\cite{Inomata:2020lmk}.  This is in contrast to the production of gravitational waves during the fragmentation of the condensate, as typically the condensate is a small fraction of the initial total energy \cite{Kusenko:2008zm,Kusenko:2009cv}.  Our proposed test is also complimentary to constraints on the ratio of scalar to isocurvature perturbations from D-term inflation~\cite{Enqvist:1999hv} and those from the backreaction of the Affleck-Dine potential on the inflaton potential, which can impact the cosmic microwave background~\cite{Marsh:2011ud}.
\par 
In this letter, we first argue that Affleck-Dine scenarios generically have this epoch of early matter domination, and secondly, the Q-ball decay rate is sufficiently fast to enhance the gravitational wave spectrum. In particular, the sudden transition avoids the suppression that occurs in a gradual transition like Moduli decay \cite{Inomata:2019zqy}.  Analytical arguments and simulations show that the Q-ball mass distribution is sharply peaked~\cite{Kusenko:1997si,Kasuya:1999wu,Multamaki:1999an,Hiramatsu:2010dx}.  Secondly, the charge quanta inside the Q-balls decay to fermions.  Q-balls decay when the decay rate per unit charge is larger than the Hubble parameter.  We show below that this is suppressed by a surface area to volume factor, and therefore the decay rate per unit charge accelerates as the Q-ball decays, similar to black hole decay.  

Furthermore, avoiding an overabundance of gravitinos results in a gravitational wave spectrum that is at sufficiently low frequencies to be observed. Finally, although in this work we make no statistical claims, we present several points in parameter space where the Q-balls are sufficiently long lived to dominate the energy density and produce a detectable gravitational wave signal. Thus, if such a signal is observed, we can narrow the cause down to two known scenarios - an early period of Q-ball domination, which is likely a consequence of Affleck-Dine baryogenesis, or an early period of light primordial black hole domination~\cite{Inomata:2020lmk,Domenech:2020ssp,Domenech:2021wkk}.  

\par 


{\bf Q-ball Induced Early Matter Domination:} during inflation, the field $\Phi$ acquires a vacuum expectation value when averaged over super-horizon scales~\cite{Bunch:1978yq,Linde:1982uu,Affleck:1984fy,Lee:1987qc,Starobinsky:1994bd}.  At the end of inflation, it relaxes towards its equilibrium value as the field fragments to form Q-balls~\cite{Kusenko:1997si,Dine:2003ax}.   During the relaxation process, a charge excess is produced as the field VEV follows a curving path in field space, which is biased either clockwise or counter-clockwise by the small CP-violating operator.  However, as a higher dimensional operator, it will also be sensitive to the initial post-inflationary VEV, which is subject to random fluctuations during inflation.  Consequently, some Hubble patches will have an excess of $Q$ charge while other have an excess of $\bar{Q}$ charge.  Therefore, there are symmetric and asymmetric components to the initial Q-ball density. 

After fragmentation, most of the condensate's initial energy is contained in Q-balls rather than individual particles, particularly if the couplings between the scalar field and fermions is small~\cite{Kusenko:1997si,Kasuya:1999wu}.  If the asymmetric component is small (as is expected due to the smallness of the observed baryon asymmetry), the symmetric component must then be large.  We parameterize the asymmetric component of the Q-ball energy density 
\begin{equation}
    r = \frac{n_{\bar{Q}}- n_{Q}}{n_{\bar{Q}}+n_Q}  \ ,
\end{equation}
and we expect $r$ to be within an order of magnitude of the baryon asymmetry.  (This can also be understood as a consequence of a highly elliptical orbit during relaxation, which simulations connect to a large symmetric component~\cite{Hiramatsu:2010dx}.)

In the thin wall regime, the vacuum expectation value inside the Q-ball can be found by minimizing $V(\Phi^2) \slash \Phi^2$ where $\Phi$ parameterizes the flat direction in the Affleck-Dine potential. The energy per unit charge of a single Q-ball is given by
\begin{align}
    \omega = \sqrt{ \dfrac{2V(v)}{ v^2}},
\end{align}
where $v$ is the VEV inside the Q-ball (we discuss specific potentials in Appendix~\ref{ap:potentials}.)  The total initial energy in Q-balls after fragmentation is
\begin{equation}
    \rho _Q = Q_0 \omega n_0
\end{equation}
where $n_0\sim N_Q H_0^3$ is the initial number density of the Q-balls and $Q_0$ is their initial charge.  Simulations suggest $N_Q \sim 1000$ for gravity-mediated SUSY scenarios and $N_Q \sim \mathcal{O}(1)$ for most gauge-mediated scenarios, if higher dimensional operators are negligible.  However, in this scenario the resulting Q-balls are in the thick wall (as opposed to thin wall) regime~\cite{Multamaki:1999an}.  Although we focus on thin wall Q-balls in this work, we note that scaling arguments suggest thick wall Q-balls are longer lived and thus also can induce early matter dominated epochs (see the supplementary material \ref{ap:potentials}).

From this, it is straightforward to derive an expression for the initial charge of the Q-balls in terms of the initial baryon asymmetry $Y_{B0}$, $r$, and the reheating temperature $T_0$,
\begin{equation}
    Q_0 = \frac{3 Y_{B0} M_{\rm Pl}^3}{800 \sqrt{5} \pi ^{5/2} g_\ast^{1\slash 2} r T_0 ^3}. \label{eq:Q0}
\end{equation}

Because the total charge in each Hubble volume is distributed into a small number of Q-balls during fragmentation, the initial Q-balls tend to have large charge; in our benchmark scenarios, the initial charges are above $10^{29}$.  Consequently, they will travel at non-relativistic speeds in the post-inflationary plasma. Then, if the Q-balls live long enough, they dominate the energy density of the universe.  We can approximate the temperature of matter-radiation equality in the limit where Q-ball decay is negligible as 
\begin{equation}
    T_{\rm eq} \sim  \frac{4 Y _{B0} \omega }{3 r} \ .
\end{equation}

Although long-lived, the Q-balls produced by the fragmentation of the Affleck-Dine condensate are not stable since their conserved charge must be transferred to Standard Model particles.  The sfermions carrying the charge can decay to a quark (or lepton) and neutralino or chargino.  Expressions for the relevant coupling can be found in Ref.~\cite{Rosiek:1995kg}, although we will parameterize the vertex in terms of an effective Yukawa coupling $y_{\rm eff}$.

This decay happens only at the surface of the Q-ball, for one of two reasons.  First, if the VEV of the squark or slepton fields inside the Q-ball (which carry the charge) is significantly larger then the energy per unit charge $\omega$, then the large induced fermion masses forbid the decay inside the Q-ball.  The induced masses of the Standard Model fermions have magnitude $gv$, where $g= g_3$ for quarks if the Q-ball is made of squarks and $g = g_2$ for leptons if the Q-balls is made of sleptons~\cite{Kusenko:2004yw}.  Therefore, if $gv \gtrsim \omega$, the decay occurs only at the surface of the Q-ball, where the VEV drops to zero.

Alternatively, if the decay is not forbidden, decays in the interior of the Q-ball rapidly fill up the Fermi sea. Thereafter, the Q-ball quanta decay only at the surface as long as the diffusion time, $t_D \sim 3 R^2/\lambda$, is sufficiently long.  The mean free path is $\lambda \sim 1/{\sigma _{\psi \phi} n}$, where number density $n = 3 Q_0/ (4 \pi R^3)$ refers to the density of scalar quanta inside the Q-ball.  The diffusion time is shortest for the highest momentum, which is $\sim \omega \slash 2$ when the decay is energetically forbidden.  The diffusion time can then be approximated using the scattering cross section $\sigma_{\psi \phi} \sim g^4_i \slash (\omega \slash 2)^2$, where $g_i \in (g_Y,g_2,g_3)$ depending on the Standard Model fermion and sfermion involved.  For the benchmark points presented below, decays inside the Q-ball are suppressed for the first reason.  

Regardless of the reason, the Q-ball evaporation rate is suppressed by the ratio of the surface area to the volume. Because the radius scales as $Q^{1\slash 3} $, the decay rate then scales as $Q^{2\slash 3} $ instead of as $Q$.  A Q-ball decays once the decay rate per unit charge is larger than the Hubble parameter.  When decays occur only on the surface, $\Gamma_{\rm Q-ball} \slash Q \propto Q^{-1 \slash 3}$, which means that it accelerates as the Q-ball shrinks.  Therefore, Q-ball decay is an  effectively instantaneous decay, similar to  black holes \cite{Inomata:2019ivs,Inomata:2020lmk}. 

The charge depletion per unit time per unit area of a Q ball obeys the equation \cite{Cohen:1986ct}
\begin{equation}
\frac{dQ}{dt\, dA}   = \frac{y_{\rm eff} v \omega ^2}{64 \pi}
\end{equation}
where $v$ is the field value of the condensate and $y_{\rm eff}$ is the effective Yukawa coupling, accounting for mixing angles.  In the thin wall limit, 
\begin{equation}
R = \left( \dfrac{3Q}{4 \pi \omega v^2} \right)^{1 \slash 3}, \label{eq:R}
\end{equation}
which gives 
\begin{equation}
\frac{\Gamma_{\rm Q-ball}}{Q} = \frac{y_{\rm eff} v \omega ^2 Q^{-1/3}}{16} \left( \frac{3}{4 \pi \omega v^2} \right)^{2 \slash 3}  . 
\end{equation}
For the Q-balls to decay after dominating the energy density, we must require $\Gamma_{\rm Q-ball} \slash HQ \big|_{T=T_{\rm eq}} \ll 1 $. 
Approximating the left side at $Q = Q_0$, we find the condition
\begin{equation}
    \frac{0.178 y_{\rm eff} r^{7/3}T_0}{  Y_{B0} ^{7/3} \omega ^{2/3} ( g_\ast v)^{1/3}}\left(\frac{1000}{N_Q} \right)^{1/3} \ll 1  \ .
\end{equation}
The large symmetric component $r \sim Y_{B0}$ is vital due to the $Y_{B0}^{- 7 \slash 3}$ factor, which would otherwise make this condition difficult to satisfy.  As expected, this prefers small Yukawa couplings, which result in long-lived Q-balls.

We emphasize that for our numerical analysis, we solved the differential equation $dQ \slash dt = -\Gamma_{\rm Q-Ball}(Q,T)$.  We also note that the decay of the Q - 
 balls dilutes the initial baryon asymmetry, resulting in a final asymmetry
\begin{equation}
    Y_{B} = Y_{B0} \left( 1+ \frac{4Y_{B0}}{3 r} T_{\rm dec} \right)^{-3/4},
\end{equation}
where $T_{\rm dec}$ is the temperature at which the Q-balls decay and $Y_B=8.59\times 10^{-11}$ as given by Planck \cite{Aghanim:2018eyx}. 

Because the Q-ball mass fraction is sharply peaked at a single value and the decay is effectively instantaneous, the scale factor and Hubble approximately obey step function solutions
\begin{equation}
    \frac{a(\eta )}{a(\eta _R) } = \left\{ \begin{array}{cc} \left( \frac{\eta}{\eta _R} \right)^2 \\ 2 \frac{\eta }{\eta _R} - 1 \end{array} \right. \ , \quad H(\eta) = \left\{  \begin{array}{cc}
        \frac{2}{\eta} & (\eta \leq \eta _R)  \\
        \frac{1}{\eta - \eta _R/2} & (\eta > \eta _R)  
    \end{array} \right.
\end{equation}
where $\eta$ is the conformal time; $\eta _R$ is specifically the conformal time at which radiation domination recommences.


{\bf Gravitational waves:} we assume inflation generates a primordial scalar power spectrum of the form
\begin{equation}
    {\cal P} _\zeta (k) = \Theta (k_{\rm inf} - k) A_s \left( \frac{k}{k_\ast} \right)^{n_s-1} \label{eq:power} 
\end{equation}
for some cutoff scale $k_{\rm inf}$, $n_s$ being the spectral tilt, $k_\ast$ being the pivot scale and $A_s$ being the amplitude at the pivot scale. We take typical values of $A_s=2.1\times 10^{-9} $, $n_s =0.97$, $k_\ast = 0.05 \ {\rm Mpc} ^{-1} $~\cite{Aghanim:2018eyx}. 

\par 
Scalar perturbations grow with the scale factor during any matter domination epoch, including the Q-ball dominated epoch, which can in turn induce gravitational waves \cite{Inomata:2019ivs,Inomata:2019zqy}.  Our analysis of the induced gravitational wave signal follows \cite{Inomata:2019zqy} and therefore we similarly work within the conformal Newtonian gauge and assume Gaussian curvature perturbations.

If matter domination is sufficiently long, then perturbations at small scales can enter the non-linear regime where a linear analysis is insufficient.  Such non-linearities become important at scales $k_{\rm NL} \sim 470/\eta _R$, where $\eta_R$ is the conformal time at which the Q-ball-caused matter domination era abruptly ended.  In this work, we neglect the non-linear regime and therefore restrict ourselves to points in parameter space at which the maximum comoving mode enhanced by early matter domination satisfies $k_{\rm max} \lesssim 470/\eta _R$.  We note that there may still be detectable gravitational wave signals in the parameter space where this is not satisfied, although we leave the analysis of the non-linear regime to future work.
\par

Using the step function approximations given above, the power spectrum of gravitational waves at conformal time $\eta$ is \cite{Kohri:2018awv}
\begin{equation}
    \Omega _{\rm GW} (\eta , k) = \frac{1}{24} \left( \frac{k}{a (\eta )H(\eta )} \right)^2 \overline{{\cal P } _h (\eta ,k)}
\end{equation}
where the time averaged power spectrum of the induced gravitational waves is related to the scalar (curvature) power spectrum by
\begin{eqnarray}
\overline{{\cal P } _h (\eta ,k)} = 4 \int _0 ^\infty dv && \int _{|1-v|}^{1+v} du \left( \frac{4 v^2-(1+v^2 - u ^2)^2}{4 v u} \right) ^2 \nonumber \\ && \times \overline{I^2(u,v,k,\eta,\eta _R} {\cal P} _\zeta (uk) {\cal P} _\zeta (vk) \ .
\end{eqnarray}
In the above, the time dependence of the gravitational waves is
\begin{equation}
    I(u,v,k,\eta,\eta _R) = \int _0 ^{k \eta} d (\overline{k \eta}) \frac{a (\bar{\eta})}{a(\eta)} k G_k (\eta , \bar{\eta} ) f(u,v, \overline{k \eta} , k \eta _R)
\end{equation}
where the Greens function is the solution of the equation
\begin{equation}
    \frac{\partial ^2 G(\eta , \bar{\eta})}{\partial \eta ^2} + \left( k^2 - \frac{1}{a}\frac{\partial ^2 a}{\partial \eta ^2} \right)G(\eta , \bar{\eta})  = \delta (\eta - \bar{\eta } ) 
\end{equation}
and the source function has the form
\begin{eqnarray}
f(u,v, \overline{k \eta} , k \eta _R) &=& \frac{3}{25(1+w)} \left( 2(5+3 w) \Phi (u \overline{k \eta} ) \Phi (v \overline{ k \eta})  \right.  \nonumber \\ 
&& \left. + 4 H^{-1} \frac{\partial}{\partial \eta } \left(  \Phi (u \overline{k \eta} ) \Phi (v \overline{ k \eta}) \right) \right. \nonumber \\ && \left. +4 H^{-2} \frac{\partial}{\partial \eta }   \Phi (u \overline{k \eta} )\frac{\partial}{\partial \eta }   \Phi (v \overline{k \eta} ) \right) \ .
\end{eqnarray}
In these equations $w$ is the equation of state parameter and $\Phi$ is the transfer function of the gravitational potential, which obeys the evolution equation \cite{Mukhanov:2005sc} 
\begin{equation}
    \frac{\partial ^2 \Phi}{\partial \eta ^2} + 3(1+w) H \frac{\partial \Phi }{\partial \eta } + w k^2 \Phi = 0 \ .
\end{equation}
For a sufficiently quick transition from matter to radiation domination, we can use the analytic formulae for the gravitational wave power spectrum in Ref. \cite{Inomata:2019ivs} which we give in the supplementary material \ref{ap:GW}. 

\begin{figure}
    \centering
    \includegraphics[width=0.48\textwidth]{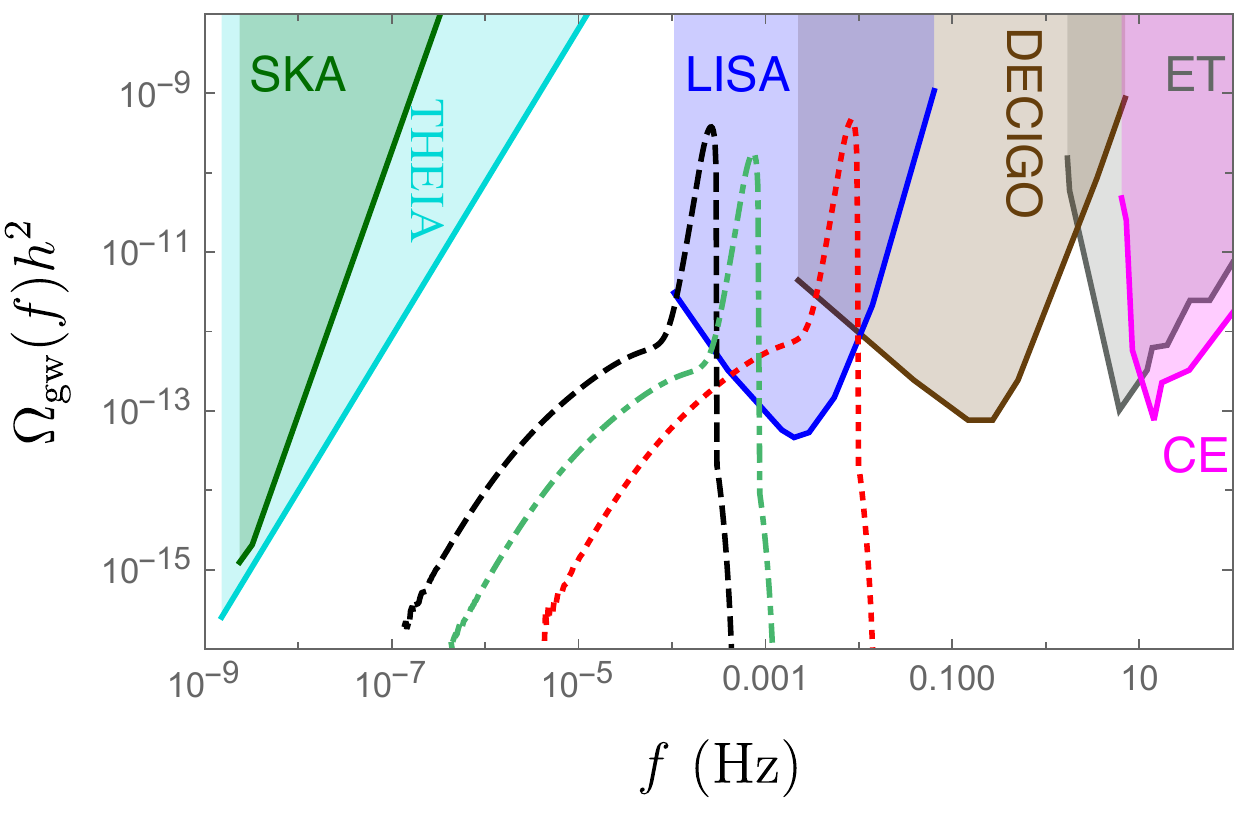}
    \caption{
 the gravitational wave signal for three benchmark scenarios, which have effective Yukawa couplings similar to the Standard Model bottom quark (red, dotted), up quark (olive, dot-dashed), and electron (black, dashed).  These clearly produce signals within the reach of future experiments, which were taken from Ref. \cite{Moore:2014lga,Kawamura:2020pcg}  for DECIGO with 3 units and an observation time of 1 year, Ref. \cite{Caprini:2019egz} for LISA with an observation time of 4 years, Ref. \cite{Boehm:2017wie,Garcia-Bellido:2021zgu} for THEIA with an observation time of 20 years, Ref. \cite{Maggiore:2019uih,Moore:2014lga,Inomata:2018epa} for Einstein Telescope with an observation time of one year, Ref. \cite{Reitze:2019iox} for the Cosmic Explorer and Ref. \cite{Janssen:2014dka} for SKA.}
    \label{fig:my_label}
\end{figure}

This rapid transition is necessary to produce the sharp peak through the ``poltergeist'' mechanism~\cite{Inomata:2020lmk}.  During the early matter domination epoch, density perturbations in non-relativistic Q-ball modes grow and form overdensities. The Q-ball decay, which is rapid as compared to the Hubble time, converts these overdensities into relativistically moving sound waves, which serve as sources of  gravitational waves. Gravitational waves exhibit a rapidly growing resonance mode which is amplified by interactions with a sound wave comoving at a certain angle~\cite{Ananda:2006af,Inomata:2019ivs,Inomata:2020lmk}.  This resonance results in a dramatic enhancement at a certain frequency, as can be seen in our Fig.~\ref{fig:my_label}.  It is important that the transition to radiation domination is rapid, because otherwise the overdensities dissolve gradually and do not result in any relativistically moving modes.


{\bf Results:} we present the gravitational wave signal for three sample points in parameter space in Fig.~\ref{fig:my_label}.  These were chosen to have Yukawa couplings similar in size to those in the Standard Model; the precise values of the parameters are given in Table~\ref{tab:my_label}.  To retain generality, we specify the VEV $v$ and energy density per charge $\omega$ of the Q-balls, instead of specializing to a particular potential.  Gravity and gauge-mediation models which produce Q-balls with these properties are discussed in the supplementary material \ref{ap:potentials}.

Calculated quantities, such as the equality and decay temperatures, for these benchmark points are given in Table~\ref{tab:my_label2}.  We note that since $\omega$ is within one order of magnitude of $T_0$, the temperature at which Q-balls are produced, it is self-consistent to neglect finite temperature corrections to $\omega$, which are induced via loop corrections.  The fragmentation of the Q-ball condensate determines the initial charge as given by Eq.\ \eqref{eq:Q0}, which produces the scaling $Q \sim 10^{34} (Y_{B0} \slash r) \cdot (10^6 \, \mathrm{GeV} \slash T_0)^3$.  The resulting values are within the typical range (see, e.g., \cite{Kasuya:2000sc}).

The observable range is controlled by the proposed frequency sensitivity of future gravitational wave detectors, with higher frequencies probing higher temperatures. At present, the highest frequency gravitational wave detectors with enough sensitivity are the Cosmic Explorer \cite{Reitze:2019iox} and the Einstein Telescope \cite{Punturo:2010zz} although higher frequency proposals are a promising work in progress \cite{Aggarwal:2020olq}.  We see that LISA has particularly good coverage of our expected signal.

There is a modest trend for points with smaller Yukawa couplings to decay later and therefore to have lower frequency peaks.  For the signal to be observable, the Q-balls must decay when the temperature falls in the range $20 \ {\rm GeV} < T_{\rm dec} < 2 \times 10^7 \ {\rm GeV} $.
While the upper bound is frequently satisfied even for large reheating temperatures, a low reheating temperature is often preferred to avoid overproduction of gravitinos, although the exact bound on the reheating temperature depends on the mass of the gravitino \cite{Khlopov:1984pf,Pradler:2006hh,Kawasaki:2008qe,Arya:2016fnf,Kawasaki:2017bqm,Eberl:2020fml}. We have imposed $T_R<10^7$ GeV for all benchmarks.

\begin{table*}[t]
    \centering
    \begin{tabular}{|c|c|c|c|c|c|c|}
    \hline 
     $\omega$ (GeV) & $v$ (GeV)  & $Y_{\bar{B}} $ & $r $ & $T_{0}$ (GeV)  & $N_Q$ & $y_{\rm eff}$  \\ \hline
$6.66 \times 10^{5}$   & $3.80 \times 10^{10}$    & $1.11 \times 10^{-8} $ & $1.56 \times 10^{-8} $ & $4.59\times 10^6$ & $ 1000$ & 0.024   \\ 
$8.45 \times 10^5$ & $1.92 \times 10^{9}$ & $1.36\times 10^{-8} $ & $2.76 \times 10^{-7}$ & $8.04 \times 10^6$ & $1000$ & $1.4 \times 10^{-5}$ \\
$9.95\times 10^5$ & $7.21 \times 10^9$ & $2.10\times 10^{-8} $ & $1.38 \times 10^{-6}$ & $3.56 \times 10^6$ & $1000$ &$2.9 \times 10^{-6}$ \\ \hline 
    \end{tabular}
    \caption{Parameters used in our three benchmark points in Fig.~\ref{fig:my_label}.  In addition to the Q-ball parameters $\omega$ and $v$, $Y_B$ is the initial charge asymmetry after fragmentation which occurs at temperature $T_0$, $N_Q$ is the average initial number of Q-balls per Hubble volume after fragmentation, and $r \sim Y_B$ is the ratio of the asymmetric component. Note that the Yukawa couplings are equal to that of the Standard Model bottom quark, up quark, and electron in the top, middle, and bottom rows.  Additionally, we have taken $g_*=106$ in our analysis.
    }
    \label{tab:my_label}
\end{table*}

\begin{table}[t]
    \centering
    \begin{tabular}{|c|c|c|}
    \hline 
$ Q_0$ & $T_{\rm eq}$ (GeV)& $T_{\rm dec}$ (GeV) \\ \hline
 $1.14 \times 10^{31}$  & $6.34 \times 10^5$ & $1368$  \\
  $1.47 \times 10^{29}$ & $55520$ & $138$   \\ 
  $5.18 \times 10^{29}$  & $20050$ & $458$ \\ \hline
    \end{tabular}
    \caption{Calculated quantities for the three benchmark points in \ref{tab:my_label}.  $T_{\rm eq}$ is the temperature of Q-ball-radiation equality and $T_{\rm dec}$ is the temperature of Q-ball decay. }
    \label{tab:my_label2}
\end{table}


{\bf Conclusions:} Affleck-Dine baryogenesis is consistent with a range of inflationary and dark matter models (see, e.g., \cite{Blinov:2021mdk,Barrie:2021mwi,Kawasaki:2020xyf,Cline:2019fxx,Yamada:2015xyr,Harigaya:2014tla,Garcia:2013bha,Kusenko:1998yi}), as a high-scale phenomena, it is difficult to directly test.  We have shown that a broad class of Affleck-Dine models produce a detectable gravitational wave signal within the range of the Einstein Telescope and/or DECIGO.  Such signals are a consequence of the sudden end of an early matter-domination epoch, which occurs if the Q-balls from the fragmentation of the Affleck-Dine condensate are sufficiently long-lived.  A low reheating temperature, motivated by the gravitino problem, ensures a signal within the observable frequency range, but we find that this is not a requirement.  Thus, if a signal is observed, the cause is one of two known scenarios- an early period of Q-ball domination, which is a natural outcome of Affleck-Dine baryogenesis, or an early period of light primordial black hole domination~\cite{Inomata:2020lmk}.

 \vspace{0.4in}
\section*{Acknowledgments} 
We thank Kazunori Kohri for useful discussions.  The works of AK and GW were supported by World Premier International Research Center Initiative (WPI), MEXT, Japan. A.K. was supported the U.S. Department of Energy (DOE) grant No. DE-SC0009937 and by Japan Society for the Promotion of Science (JSPS) KAKENHI grant No.
JP20H05853. 

\bibliography{references}

\begin{thebibliography}{62}%
\makeatletter
\providecommand \@ifxundefined [1]{%
 \@ifx{#1\undefined}
}%
\providecommand \@ifnum [1]{%
 \ifnum #1\expandafter \@firstoftwo
 \else \expandafter \@secondoftwo
 \fi
}%
\providecommand \@ifx [1]{%
 \ifx #1\expandafter \@firstoftwo
 \else \expandafter \@secondoftwo
 \fi
}%
\providecommand \natexlab [1]{#1}%
\providecommand \enquote  [1]{``#1''}%
\providecommand \bibnamefont  [1]{#1}%
\providecommand \bibfnamefont [1]{#1}%
\providecommand \citenamefont [1]{#1}%
\providecommand \href@noop [0]{\@secondoftwo}%
\providecommand \href [0]{\begingroup \@sanitize@url \@href}%
\providecommand \@href[1]{\@@startlink{#1}\@@href}%
\providecommand \@@href[1]{\endgroup#1\@@endlink}%
\providecommand \@sanitize@url [0]{\catcode `\\12\catcode `\$12\catcode
  `\&12\catcode `\#12\catcode `\^12\catcode `\_12\catcode `\%12\relax}%
\providecommand \@@startlink[1]{}%
\providecommand \@@endlink[0]{}%
\providecommand \url  [0]{\begingroup\@sanitize@url \@url }%
\providecommand \@url [1]{\endgroup\@href {#1}{\urlprefix }}%
\providecommand \urlprefix  [0]{URL }%
\providecommand \Eprint [0]{\href }%
\providecommand \doibase [0]{http://dx.doi.org/}%
\providecommand \selectlanguage [0]{\@gobble}%
\providecommand \bibinfo  [0]{\@secondoftwo}%
\providecommand \bibfield  [0]{\@secondoftwo}%
\providecommand \translation [1]{[#1]}%
\providecommand \BibitemOpen [0]{}%
\providecommand \bibitemStop [0]{}%
\providecommand \bibitemNoStop [0]{.\EOS\space}%
\providecommand \EOS [0]{\spacefactor3000\relax}%
\providecommand \BibitemShut  [1]{\csname bibitem#1\endcsname}%
\let\auto@bib@innerbib\@empty
\bibitem [{\citenamefont {Sakharov}(1967)}]{Sakharov:1967dj}%
  \BibitemOpen
  \bibfield  {author} {\bibinfo {author} {\bibfnamefont {A.~D.}\ \bibnamefont
  {Sakharov}},\ }\href {\doibase 10.1070/PU1991v034n05ABEH002497} {\bibfield
  {journal} {\bibinfo  {journal} {Pisma Zh. Eksp. Teor. Fiz.}\ }\textbf
  {\bibinfo {volume} {5}},\ \bibinfo {pages} {32} (\bibinfo {year}
  {1967})}\BibitemShut {NoStop}%
\bibitem [{\citenamefont {Shaposhnikov}(1987)}]{Shaposhnikov:1987tw}%
  \BibitemOpen
  \bibfield  {author} {\bibinfo {author} {\bibfnamefont {M.~E.}\ \bibnamefont
  {Shaposhnikov}},\ }\href {\doibase 10.1016/0550-3213(87)90127-1} {\bibfield
  {journal} {\bibinfo  {journal} {Nucl. Phys. B}\ }\textbf {\bibinfo {volume}
  {287}},\ \bibinfo {pages} {757} (\bibinfo {year} {1987})}\BibitemShut
  {NoStop}%
\bibitem [{\citenamefont {Dine}\ and\ \citenamefont
  {Kusenko}(2003)}]{Dine:2003ax}%
  \BibitemOpen
  \bibfield  {author} {\bibinfo {author} {\bibfnamefont {M.}~\bibnamefont
  {Dine}}\ and\ \bibinfo {author} {\bibfnamefont {A.}~\bibnamefont {Kusenko}},\
  }\href {\doibase 10.1103/RevModPhys.76.1} {\bibfield  {journal} {\bibinfo
  {journal} {Rev. Mod. Phys.}\ }\textbf {\bibinfo {volume} {76}},\ \bibinfo
  {pages} {1} (\bibinfo {year} {2003})},\ \Eprint
  {http://arxiv.org/abs/hep-ph/0303065} {arXiv:hep-ph/0303065} \BibitemShut
  {NoStop}%
\bibitem [{\citenamefont {Affleck}\ and\ \citenamefont
  {Dine}(1985)}]{Affleck:1984fy}%
  \BibitemOpen
  \bibfield  {author} {\bibinfo {author} {\bibfnamefont {I.}~\bibnamefont
  {Affleck}}\ and\ \bibinfo {author} {\bibfnamefont {M.}~\bibnamefont {Dine}},\
  }\href {\doibase 10.1016/0550-3213(85)90021-5} {\bibfield  {journal}
  {\bibinfo  {journal} {Nucl. Phys. B}\ }\textbf {\bibinfo {volume} {249}},\
  \bibinfo {pages} {361} (\bibinfo {year} {1985})}\BibitemShut {NoStop}%
\bibitem [{\citenamefont {Dine}\ \emph {et~al.}(1996)\citenamefont {Dine},
  \citenamefont {Randall},\ and\ \citenamefont {Thomas}}]{Dine:1995kz}%
  \BibitemOpen
  \bibfield  {author} {\bibinfo {author} {\bibfnamefont {M.}~\bibnamefont
  {Dine}}, \bibinfo {author} {\bibfnamefont {L.}~\bibnamefont {Randall}}, \
  and\ \bibinfo {author} {\bibfnamefont {S.~D.}\ \bibnamefont {Thomas}},\
  }\href {\doibase 10.1016/0550-3213(95)00538-2} {\bibfield  {journal}
  {\bibinfo  {journal} {Nucl. Phys. B}\ }\textbf {\bibinfo {volume} {458}},\
  \bibinfo {pages} {291} (\bibinfo {year} {1996})},\ \Eprint
  {http://arxiv.org/abs/hep-ph/9507453} {arXiv:hep-ph/9507453} \BibitemShut
  {NoStop}%
\bibitem [{\citenamefont {Allahverdi}\ and\ \citenamefont
  {Mazumdar}(2012)}]{Allahverdi:2012ju}%
  \BibitemOpen
  \bibfield  {author} {\bibinfo {author} {\bibfnamefont {R.}~\bibnamefont
  {Allahverdi}}\ and\ \bibinfo {author} {\bibfnamefont {A.}~\bibnamefont
  {Mazumdar}},\ }\href {\doibase 10.1088/1367-2630/14/12/125013} {\bibfield
  {journal} {\bibinfo  {journal} {New J. Phys.}\ }\textbf {\bibinfo {volume}
  {14}},\ \bibinfo {pages} {125013} (\bibinfo {year} {2012})}\BibitemShut
  {NoStop}%
\bibitem [{\citenamefont {Gherghetta}\ \emph {et~al.}(1996)\citenamefont
  {Gherghetta}, \citenamefont {Kolda},\ and\ \citenamefont
  {Martin}}]{Gherghetta:1995dv}%
  \BibitemOpen
  \bibfield  {author} {\bibinfo {author} {\bibfnamefont {T.}~\bibnamefont
  {Gherghetta}}, \bibinfo {author} {\bibfnamefont {C.~F.}\ \bibnamefont
  {Kolda}}, \ and\ \bibinfo {author} {\bibfnamefont {S.~P.}\ \bibnamefont
  {Martin}},\ }\href {\doibase 10.1016/0550-3213(96)00095-8} {\bibfield
  {journal} {\bibinfo  {journal} {Nucl. Phys. B}\ }\textbf {\bibinfo {volume}
  {468}},\ \bibinfo {pages} {37} (\bibinfo {year} {1996})},\ \Eprint
  {http://arxiv.org/abs/hep-ph/9510370} {arXiv:hep-ph/9510370} \BibitemShut
  {NoStop}%
\bibitem [{\citenamefont {Kusenko}\ and\ \citenamefont
  {Shaposhnikov}(1998)}]{Kusenko:1997si}%
  \BibitemOpen
  \bibfield  {author} {\bibinfo {author} {\bibfnamefont {A.}~\bibnamefont
  {Kusenko}}\ and\ \bibinfo {author} {\bibfnamefont {M.~E.}\ \bibnamefont
  {Shaposhnikov}},\ }\href {\doibase 10.1016/S0370-2693(97)01375-0} {\bibfield
  {journal} {\bibinfo  {journal} {Phys. Lett. B}\ }\textbf {\bibinfo {volume}
  {418}},\ \bibinfo {pages} {46} (\bibinfo {year} {1998})},\ \Eprint
  {http://arxiv.org/abs/hep-ph/9709492} {arXiv:hep-ph/9709492} \BibitemShut
  {NoStop}%
\bibitem [{\citenamefont {Rosen}(1968)}]{Rosen:1968mfz}%
  \BibitemOpen
  \bibfield  {author} {\bibinfo {author} {\bibfnamefont {G.}~\bibnamefont
  {Rosen}},\ }\href {\doibase 10.1063/1.1664693} {\bibfield  {journal}
  {\bibinfo  {journal} {J. Math. Phys.}\ }\textbf {\bibinfo {volume} {9}},\
  \bibinfo {pages} {996} (\bibinfo {year} {1968})}\BibitemShut {NoStop}%
\bibitem [{\citenamefont {Friedberg}\ \emph {et~al.}(1976)\citenamefont
  {Friedberg}, \citenamefont {Lee},\ and\ \citenamefont
  {Sirlin}}]{Friedberg:1976me}%
  \BibitemOpen
  \bibfield  {author} {\bibinfo {author} {\bibfnamefont {R.}~\bibnamefont
  {Friedberg}}, \bibinfo {author} {\bibfnamefont {T.~D.}\ \bibnamefont {Lee}},
  \ and\ \bibinfo {author} {\bibfnamefont {A.}~\bibnamefont {Sirlin}},\ }\href
  {\doibase 10.1103/PhysRevD.13.2739} {\bibfield  {journal} {\bibinfo
  {journal} {Phys. Rev. D}\ }\textbf {\bibinfo {volume} {13}},\ \bibinfo
  {pages} {2739} (\bibinfo {year} {1976})}\BibitemShut {NoStop}%
\bibitem [{\citenamefont {Coleman}(1985)}]{Coleman:1985ki}%
  \BibitemOpen
  \bibfield  {author} {\bibinfo {author} {\bibfnamefont {S.~R.}\ \bibnamefont
  {Coleman}},\ }\href {\doibase 10.1016/0550-3213(85)90286-X,
  10.1016/0550-3213(86)90520-1} {\bibfield  {journal} {\bibinfo  {journal}
  {Nucl. Phys.}\ }\textbf {\bibinfo {volume} {B262}},\ \bibinfo {pages} {263}
  (\bibinfo {year} {1985})},\ \bibinfo {note} {[Erratum: Nucl.
  Phys.B269,744(1986)]}\BibitemShut {NoStop}%
\bibitem [{\citenamefont {Kusenko}(1997{\natexlab{a}})}]{Kusenko:1997ad}%
  \BibitemOpen
  \bibfield  {author} {\bibinfo {author} {\bibfnamefont {A.}~\bibnamefont
  {Kusenko}},\ }\href {\doibase 10.1016/S0370-2693(97)00582-0} {\bibfield
  {journal} {\bibinfo  {journal} {Phys. Lett. B}\ }\textbf {\bibinfo {volume}
  {404}},\ \bibinfo {pages} {285} (\bibinfo {year} {1997}{\natexlab{a}})},\
  \Eprint {http://arxiv.org/abs/hep-th/9704073} {arXiv:hep-th/9704073}
  \BibitemShut {NoStop}%
\bibitem [{\citenamefont {Kusenko}(1997{\natexlab{b}})}]{Kusenko:1997zq}%
  \BibitemOpen
  \bibfield  {author} {\bibinfo {author} {\bibfnamefont {A.}~\bibnamefont
  {Kusenko}},\ }\href {\doibase 10.1016/S0370-2693(97)00584-4} {\bibfield
  {journal} {\bibinfo  {journal} {Phys. Lett. B}\ }\textbf {\bibinfo {volume}
  {405}},\ \bibinfo {pages} {108} (\bibinfo {year} {1997}{\natexlab{b}})},\
  \Eprint {http://arxiv.org/abs/hep-ph/9704273} {arXiv:hep-ph/9704273}
  \BibitemShut {NoStop}%
\bibitem [{\citenamefont {Inomata}\ \emph {et~al.}(2020)\citenamefont
  {Inomata}, \citenamefont {Kawasaki}, \citenamefont {Mukaida}, \citenamefont
  {Terada},\ and\ \citenamefont {Yanagida}}]{Inomata:2020lmk}%
  \BibitemOpen
  \bibfield  {author} {\bibinfo {author} {\bibfnamefont {K.}~\bibnamefont
  {Inomata}}, \bibinfo {author} {\bibfnamefont {M.}~\bibnamefont {Kawasaki}},
  \bibinfo {author} {\bibfnamefont {K.}~\bibnamefont {Mukaida}}, \bibinfo
  {author} {\bibfnamefont {T.}~\bibnamefont {Terada}}, \ and\ \bibinfo {author}
  {\bibfnamefont {T.~T.}\ \bibnamefont {Yanagida}},\ }\href {\doibase
  10.1103/PhysRevD.101.123533} {\bibfield  {journal} {\bibinfo  {journal}
  {Phys. Rev. D}\ }\textbf {\bibinfo {volume} {101}},\ \bibinfo {pages}
  {123533} (\bibinfo {year} {2020})},\ \Eprint
  {http://arxiv.org/abs/2003.10455} {arXiv:2003.10455 [astro-ph.CO]}
  \BibitemShut {NoStop}%
\bibitem [{\citenamefont {Kusenko}\ and\ \citenamefont
  {Mazumdar}(2008)}]{Kusenko:2008zm}%
  \BibitemOpen
  \bibfield  {author} {\bibinfo {author} {\bibfnamefont {A.}~\bibnamefont
  {Kusenko}}\ and\ \bibinfo {author} {\bibfnamefont {A.}~\bibnamefont
  {Mazumdar}},\ }\href {\doibase 10.1103/PhysRevLett.101.211301} {\bibfield
  {journal} {\bibinfo  {journal} {Phys. Rev. Lett.}\ }\textbf {\bibinfo
  {volume} {101}},\ \bibinfo {pages} {211301} (\bibinfo {year} {2008})},\
  \Eprint {http://arxiv.org/abs/0807.4554} {arXiv:0807.4554 [astro-ph]}
  \BibitemShut {NoStop}%
\bibitem [{\citenamefont {Kusenko}\ \emph {et~al.}(2009)\citenamefont
  {Kusenko}, \citenamefont {Mazumdar},\ and\ \citenamefont
  {Multamaki}}]{Kusenko:2009cv}%
  \BibitemOpen
  \bibfield  {author} {\bibinfo {author} {\bibfnamefont {A.}~\bibnamefont
  {Kusenko}}, \bibinfo {author} {\bibfnamefont {A.}~\bibnamefont {Mazumdar}}, \
  and\ \bibinfo {author} {\bibfnamefont {T.}~\bibnamefont {Multamaki}},\ }\href
  {\doibase 10.1103/PhysRevD.79.124034} {\bibfield  {journal} {\bibinfo
  {journal} {Phys. Rev. D}\ }\textbf {\bibinfo {volume} {79}},\ \bibinfo
  {pages} {124034} (\bibinfo {year} {2009})},\ \Eprint
  {http://arxiv.org/abs/0902.2197} {arXiv:0902.2197 [astro-ph.CO]} \BibitemShut
  {NoStop}%
\bibitem [{\citenamefont {Enqvist}\ and\ \citenamefont
  {McDonald}(2000)}]{Enqvist:1999hv}%
  \BibitemOpen
  \bibfield  {author} {\bibinfo {author} {\bibfnamefont {K.}~\bibnamefont
  {Enqvist}}\ and\ \bibinfo {author} {\bibfnamefont {J.}~\bibnamefont
  {McDonald}},\ }\href {\doibase 10.1103/PhysRevD.62.043502} {\bibfield
  {journal} {\bibinfo  {journal} {Phys. Rev. D}\ }\textbf {\bibinfo {volume}
  {62}},\ \bibinfo {pages} {043502} (\bibinfo {year} {2000})},\ \Eprint
  {http://arxiv.org/abs/hep-ph/9912478} {arXiv:hep-ph/9912478} \BibitemShut
  {NoStop}%
\bibitem [{\citenamefont {Marsh}(2012)}]{Marsh:2011ud}%
  \BibitemOpen
  \bibfield  {author} {\bibinfo {author} {\bibfnamefont {D.}~\bibnamefont
  {Marsh}},\ }\href {\doibase 10.1007/JHEP05(2012)041} {\bibfield  {journal}
  {\bibinfo  {journal} {JHEP}\ }\textbf {\bibinfo {volume} {05}},\ \bibinfo
  {pages} {041} (\bibinfo {year} {2012})},\ \Eprint
  {http://arxiv.org/abs/1108.4687} {arXiv:1108.4687 [hep-th]} \BibitemShut
  {NoStop}%
\bibitem [{\citenamefont {Inomata}\ \emph
  {et~al.}(2019{\natexlab{a}})\citenamefont {Inomata}, \citenamefont {Kohri},
  \citenamefont {Nakama},\ and\ \citenamefont {Terada}}]{Inomata:2019zqy}%
  \BibitemOpen
  \bibfield  {author} {\bibinfo {author} {\bibfnamefont {K.}~\bibnamefont
  {Inomata}}, \bibinfo {author} {\bibfnamefont {K.}~\bibnamefont {Kohri}},
  \bibinfo {author} {\bibfnamefont {T.}~\bibnamefont {Nakama}}, \ and\ \bibinfo
  {author} {\bibfnamefont {T.}~\bibnamefont {Terada}},\ }\href {\doibase
  10.1088/1475-7516/2019/10/071} {\bibfield  {journal} {\bibinfo  {journal}
  {JCAP}\ }\textbf {\bibinfo {volume} {10}},\ \bibinfo {pages} {071} (\bibinfo
  {year} {2019}{\natexlab{a}})},\ \Eprint {http://arxiv.org/abs/1904.12878}
  {arXiv:1904.12878 [astro-ph.CO]} \BibitemShut {NoStop}%
\bibitem [{\citenamefont {Kasuya}\ and\ \citenamefont
  {Kawasaki}(2000{\natexlab{a}})}]{Kasuya:1999wu}%
  \BibitemOpen
  \bibfield  {author} {\bibinfo {author} {\bibfnamefont {S.}~\bibnamefont
  {Kasuya}}\ and\ \bibinfo {author} {\bibfnamefont {M.}~\bibnamefont
  {Kawasaki}},\ }\href {\doibase 10.1103/PhysRevD.61.041301} {\bibfield
  {journal} {\bibinfo  {journal} {Phys. Rev. D}\ }\textbf {\bibinfo {volume}
  {61}},\ \bibinfo {pages} {041301} (\bibinfo {year} {2000}{\natexlab{a}})},\
  \Eprint {http://arxiv.org/abs/hep-ph/9909509} {arXiv:hep-ph/9909509}
  \BibitemShut {NoStop}%
\bibitem [{\citenamefont {Multamaki}\ and\ \citenamefont
  {Vilja}(2000)}]{Multamaki:1999an}%
  \BibitemOpen
  \bibfield  {author} {\bibinfo {author} {\bibfnamefont {T.}~\bibnamefont
  {Multamaki}}\ and\ \bibinfo {author} {\bibfnamefont {I.}~\bibnamefont
  {Vilja}},\ }\href {\doibase 10.1016/S0550-3213(99)00827-5} {\bibfield
  {journal} {\bibinfo  {journal} {Nucl. Phys. B}\ }\textbf {\bibinfo {volume}
  {574}},\ \bibinfo {pages} {130} (\bibinfo {year} {2000})},\ \Eprint
  {http://arxiv.org/abs/hep-ph/9908446} {arXiv:hep-ph/9908446} \BibitemShut
  {NoStop}%
\bibitem [{\citenamefont {Hiramatsu}\ \emph {et~al.}(2010)\citenamefont
  {Hiramatsu}, \citenamefont {Kawasaki},\ and\ \citenamefont
  {Takahashi}}]{Hiramatsu:2010dx}%
  \BibitemOpen
  \bibfield  {author} {\bibinfo {author} {\bibfnamefont {T.}~\bibnamefont
  {Hiramatsu}}, \bibinfo {author} {\bibfnamefont {M.}~\bibnamefont {Kawasaki}},
  \ and\ \bibinfo {author} {\bibfnamefont {F.}~\bibnamefont {Takahashi}},\
  }\href {\doibase 10.1088/1475-7516/2010/06/008} {\bibfield  {journal}
  {\bibinfo  {journal} {JCAP}\ }\textbf {\bibinfo {volume} {06}},\ \bibinfo
  {pages} {008} (\bibinfo {year} {2010})},\ \Eprint
  {http://arxiv.org/abs/1003.1779} {arXiv:1003.1779 [hep-ph]} \BibitemShut
  {NoStop}%
\bibitem [{\citenamefont {Dom\`enech}\ \emph
  {et~al.}(2021{\natexlab{a}})\citenamefont {Dom\`enech}, \citenamefont {Lin},\
  and\ \citenamefont {Sasaki}}]{Domenech:2020ssp}%
  \BibitemOpen
  \bibfield  {author} {\bibinfo {author} {\bibfnamefont {G.}~\bibnamefont
  {Dom\`enech}}, \bibinfo {author} {\bibfnamefont {C.}~\bibnamefont {Lin}}, \
  and\ \bibinfo {author} {\bibfnamefont {M.}~\bibnamefont {Sasaki}},\ }\href
  {\doibase 10.1088/1475-7516/2021/04/062} {\bibfield  {journal} {\bibinfo
  {journal} {JCAP}\ }\textbf {\bibinfo {volume} {04}},\ \bibinfo {pages} {062}
  (\bibinfo {year} {2021}{\natexlab{a}})},\ \Eprint
  {http://arxiv.org/abs/2012.08151} {arXiv:2012.08151 [gr-qc]} \BibitemShut
  {NoStop}%
\bibitem [{\citenamefont {Dom\`enech}\ \emph
  {et~al.}(2021{\natexlab{b}})\citenamefont {Dom\`enech}, \citenamefont
  {Takhistov},\ and\ \citenamefont {Sasaki}}]{Domenech:2021wkk}%
  \BibitemOpen
  \bibfield  {author} {\bibinfo {author} {\bibfnamefont {G.}~\bibnamefont
  {Dom\`enech}}, \bibinfo {author} {\bibfnamefont {V.}~\bibnamefont
  {Takhistov}}, \ and\ \bibinfo {author} {\bibfnamefont {M.}~\bibnamefont
  {Sasaki}},\ }\href@noop {} {\  (\bibinfo {year} {2021}{\natexlab{b}})},\
  \Eprint {http://arxiv.org/abs/2105.06816} {arXiv:2105.06816 [astro-ph.CO]}
  \BibitemShut {NoStop}%
\bibitem [{\citenamefont {Bunch}\ and\ \citenamefont
  {Davies}(1978)}]{Bunch:1978yq}%
  \BibitemOpen
  \bibfield  {author} {\bibinfo {author} {\bibfnamefont {T.~S.}\ \bibnamefont
  {Bunch}}\ and\ \bibinfo {author} {\bibfnamefont {P.~C.~W.}\ \bibnamefont
  {Davies}},\ }\href {\doibase 10.1098/rspa.1978.0060} {\bibfield  {journal}
  {\bibinfo  {journal} {Proc. Roy. Soc. Lond. A}\ }\textbf {\bibinfo {volume}
  {360}},\ \bibinfo {pages} {117} (\bibinfo {year} {1978})}\BibitemShut
  {NoStop}%
\bibitem [{\citenamefont {Linde}(1982)}]{Linde:1982uu}%
  \BibitemOpen
  \bibfield  {author} {\bibinfo {author} {\bibfnamefont {A.~D.}\ \bibnamefont
  {Linde}},\ }\href {\doibase 10.1016/0370-2693(82)90293-3} {\bibfield
  {journal} {\bibinfo  {journal} {Phys. Lett. B}\ }\textbf {\bibinfo {volume}
  {116}},\ \bibinfo {pages} {335} (\bibinfo {year} {1982})}\BibitemShut
  {NoStop}%
\bibitem [{\citenamefont {Lee}\ and\ \citenamefont
  {Weinberg}(1987)}]{Lee:1987qc}%
  \BibitemOpen
  \bibfield  {author} {\bibinfo {author} {\bibfnamefont {K.-M.}\ \bibnamefont
  {Lee}}\ and\ \bibinfo {author} {\bibfnamefont {E.~J.}\ \bibnamefont
  {Weinberg}},\ }\href {\doibase 10.1103/PhysRevD.36.1088} {\bibfield
  {journal} {\bibinfo  {journal} {Phys. Rev. D}\ }\textbf {\bibinfo {volume}
  {36}},\ \bibinfo {pages} {1088} (\bibinfo {year} {1987})}\BibitemShut
  {NoStop}%
\bibitem [{\citenamefont {Starobinsky}\ and\ \citenamefont
  {Yokoyama}(1994)}]{Starobinsky:1994bd}%
  \BibitemOpen
  \bibfield  {author} {\bibinfo {author} {\bibfnamefont {A.~A.}\ \bibnamefont
  {Starobinsky}}\ and\ \bibinfo {author} {\bibfnamefont {J.}~\bibnamefont
  {Yokoyama}},\ }\href {\doibase 10.1103/PhysRevD.50.6357} {\bibfield
  {journal} {\bibinfo  {journal} {Phys. Rev. D}\ }\textbf {\bibinfo {volume}
  {50}},\ \bibinfo {pages} {6357} (\bibinfo {year} {1994})},\ \Eprint
  {http://arxiv.org/abs/astro-ph/9407016} {arXiv:astro-ph/9407016} \BibitemShut
  {NoStop}%
\bibitem [{\citenamefont {Rosiek}(1995)}]{Rosiek:1995kg}%
  \BibitemOpen
  \bibfield  {author} {\bibinfo {author} {\bibfnamefont {J.}~\bibnamefont
  {Rosiek}},\ }\href@noop {} {\  (\bibinfo {year} {1995})},\ \Eprint
  {http://arxiv.org/abs/hep-ph/9511250} {arXiv:hep-ph/9511250} \BibitemShut
  {NoStop}%
\bibitem [{\citenamefont {Kusenko}\ \emph {et~al.}(2005)\citenamefont
  {Kusenko}, \citenamefont {Loveridge},\ and\ \citenamefont
  {Shaposhnikov}}]{Kusenko:2004yw}%
  \BibitemOpen
  \bibfield  {author} {\bibinfo {author} {\bibfnamefont {A.}~\bibnamefont
  {Kusenko}}, \bibinfo {author} {\bibfnamefont {L.}~\bibnamefont {Loveridge}},
  \ and\ \bibinfo {author} {\bibfnamefont {M.}~\bibnamefont {Shaposhnikov}},\
  }\href {\doibase 10.1103/PhysRevD.72.025015} {\bibfield  {journal} {\bibinfo
  {journal} {Phys. Rev. D}\ }\textbf {\bibinfo {volume} {72}},\ \bibinfo
  {pages} {025015} (\bibinfo {year} {2005})},\ \Eprint
  {http://arxiv.org/abs/hep-ph/0405044} {arXiv:hep-ph/0405044} \BibitemShut
  {NoStop}%
\bibitem [{\citenamefont {Inomata}\ \emph
  {et~al.}(2019{\natexlab{b}})\citenamefont {Inomata}, \citenamefont {Kohri},
  \citenamefont {Nakama},\ and\ \citenamefont {Terada}}]{Inomata:2019ivs}%
  \BibitemOpen
  \bibfield  {author} {\bibinfo {author} {\bibfnamefont {K.}~\bibnamefont
  {Inomata}}, \bibinfo {author} {\bibfnamefont {K.}~\bibnamefont {Kohri}},
  \bibinfo {author} {\bibfnamefont {T.}~\bibnamefont {Nakama}}, \ and\ \bibinfo
  {author} {\bibfnamefont {T.}~\bibnamefont {Terada}},\ }\href {\doibase
  10.1103/PhysRevD.100.043532} {\bibfield  {journal} {\bibinfo  {journal}
  {Phys. Rev. D}\ }\textbf {\bibinfo {volume} {100}},\ \bibinfo {pages}
  {043532} (\bibinfo {year} {2019}{\natexlab{b}})},\ \Eprint
  {http://arxiv.org/abs/1904.12879} {arXiv:1904.12879 [astro-ph.CO]}
  \BibitemShut {NoStop}%
\bibitem [{\citenamefont {Cohen}\ \emph {et~al.}(1986)\citenamefont {Cohen},
  \citenamefont {Coleman}, \citenamefont {Georgi},\ and\ \citenamefont
  {Manohar}}]{Cohen:1986ct}%
  \BibitemOpen
  \bibfield  {author} {\bibinfo {author} {\bibfnamefont {A.~G.}\ \bibnamefont
  {Cohen}}, \bibinfo {author} {\bibfnamefont {S.~R.}\ \bibnamefont {Coleman}},
  \bibinfo {author} {\bibfnamefont {H.}~\bibnamefont {Georgi}}, \ and\ \bibinfo
  {author} {\bibfnamefont {A.}~\bibnamefont {Manohar}},\ }\href {\doibase
  10.1016/0550-3213(86)90004-0} {\bibfield  {journal} {\bibinfo  {journal}
  {Nucl. Phys. B}\ }\textbf {\bibinfo {volume} {272}},\ \bibinfo {pages} {301}
  (\bibinfo {year} {1986})}\BibitemShut {NoStop}%
\bibitem [{\citenamefont {Aghanim}\ \emph {et~al.}(2020)\citenamefont {Aghanim}
  \emph {et~al.}}]{Aghanim:2018eyx}%
  \BibitemOpen
  \bibfield  {author} {\bibinfo {author} {\bibfnamefont {N.}~\bibnamefont
  {Aghanim}} \emph {et~al.} (\bibinfo {collaboration} {Planck}),\ }\href
  {\doibase 10.1051/0004-6361/201833910} {\bibfield  {journal} {\bibinfo
  {journal} {Astron. Astrophys.}\ }\textbf {\bibinfo {volume} {641}},\ \bibinfo
  {pages} {A6} (\bibinfo {year} {2020})},\ \Eprint
  {http://arxiv.org/abs/1807.06209} {arXiv:1807.06209 [astro-ph.CO]}
  \BibitemShut {NoStop}%
\bibitem [{\citenamefont {Kohri}\ and\ \citenamefont
  {Terada}(2018)}]{Kohri:2018awv}%
  \BibitemOpen
  \bibfield  {author} {\bibinfo {author} {\bibfnamefont {K.}~\bibnamefont
  {Kohri}}\ and\ \bibinfo {author} {\bibfnamefont {T.}~\bibnamefont {Terada}},\
  }\href {\doibase 10.1103/PhysRevD.97.123532} {\bibfield  {journal} {\bibinfo
  {journal} {Phys. Rev. D}\ }\textbf {\bibinfo {volume} {97}},\ \bibinfo
  {pages} {123532} (\bibinfo {year} {2018})},\ \Eprint
  {http://arxiv.org/abs/1804.08577} {arXiv:1804.08577 [gr-qc]} \BibitemShut
  {NoStop}%
\bibitem [{\citenamefont {Mukhanov}(2005)}]{Mukhanov:2005sc}%
  \BibitemOpen
  \bibfield  {author} {\bibinfo {author} {\bibfnamefont {V.}~\bibnamefont
  {Mukhanov}},\ }\href@noop {} {\emph {\bibinfo {title} {{Physical Foundations
  of Cosmology}}}}\ (\bibinfo  {publisher} {Cambridge University Press},\
  \bibinfo {address} {Oxford},\ \bibinfo {year} {2005})\BibitemShut {NoStop}%
\bibitem [{\citenamefont {Ananda}\ \emph {et~al.}(2007)\citenamefont {Ananda},
  \citenamefont {Clarkson},\ and\ \citenamefont {Wands}}]{Ananda:2006af}%
  \BibitemOpen
  \bibfield  {author} {\bibinfo {author} {\bibfnamefont {K.~N.}\ \bibnamefont
  {Ananda}}, \bibinfo {author} {\bibfnamefont {C.}~\bibnamefont {Clarkson}}, \
  and\ \bibinfo {author} {\bibfnamefont {D.}~\bibnamefont {Wands}},\ }\href
  {\doibase 10.1103/PhysRevD.75.123518} {\bibfield  {journal} {\bibinfo
  {journal} {Phys. Rev. D}\ }\textbf {\bibinfo {volume} {75}},\ \bibinfo
  {pages} {123518} (\bibinfo {year} {2007})},\ \Eprint
  {http://arxiv.org/abs/gr-qc/0612013} {arXiv:gr-qc/0612013} \BibitemShut
  {NoStop}%
\bibitem [{\citenamefont {Moore}\ \emph {et~al.}(2015)\citenamefont {Moore},
  \citenamefont {Cole},\ and\ \citenamefont {Berry}}]{Moore:2014lga}%
  \BibitemOpen
  \bibfield  {author} {\bibinfo {author} {\bibfnamefont {C.~J.}\ \bibnamefont
  {Moore}}, \bibinfo {author} {\bibfnamefont {R.~H.}\ \bibnamefont {Cole}}, \
  and\ \bibinfo {author} {\bibfnamefont {C.~P.~L.}\ \bibnamefont {Berry}},\
  }\href {\doibase 10.1088/0264-9381/32/1/015014} {\bibfield  {journal}
  {\bibinfo  {journal} {Class. Quant. Grav.}\ }\textbf {\bibinfo {volume}
  {32}},\ \bibinfo {pages} {015014} (\bibinfo {year} {2015})},\ \Eprint
  {http://arxiv.org/abs/1408.0740} {arXiv:1408.0740 [gr-qc]} \BibitemShut
  {NoStop}%
\bibitem [{\citenamefont {Kawamura}\ \emph {et~al.}(2020)\citenamefont
  {Kawamura} \emph {et~al.}}]{Kawamura:2020pcg}%
  \BibitemOpen
  \bibfield  {author} {\bibinfo {author} {\bibfnamefont {S.}~\bibnamefont
  {Kawamura}} \emph {et~al.},\ }\href@noop {} {\  (\bibinfo {year} {2020})},\
  \Eprint {http://arxiv.org/abs/2006.13545} {arXiv:2006.13545 [gr-qc]}
  \BibitemShut {NoStop}%
\bibitem [{\citenamefont {Caprini}\ \emph {et~al.}(2020)\citenamefont {Caprini}
  \emph {et~al.}}]{Caprini:2019egz}%
  \BibitemOpen
  \bibfield  {author} {\bibinfo {author} {\bibfnamefont {C.}~\bibnamefont
  {Caprini}} \emph {et~al.},\ }\href {\doibase 10.1088/1475-7516/2020/03/024}
  {\bibfield  {journal} {\bibinfo  {journal} {JCAP}\ }\textbf {\bibinfo
  {volume} {03}},\ \bibinfo {pages} {024} (\bibinfo {year} {2020})},\ \Eprint
  {http://arxiv.org/abs/1910.13125} {arXiv:1910.13125 [astro-ph.CO]}
  \BibitemShut {NoStop}%
\bibitem [{\citenamefont {Boehm}\ \emph {et~al.}(2017)\citenamefont {Boehm}
  \emph {et~al.}}]{Boehm:2017wie}%
  \BibitemOpen
  \bibfield  {author} {\bibinfo {author} {\bibfnamefont {C.}~\bibnamefont
  {Boehm}} \emph {et~al.} (\bibinfo {collaboration} {Theia}),\ }\href@noop {}
  {\  (\bibinfo {year} {2017})},\ \Eprint {http://arxiv.org/abs/1707.01348}
  {arXiv:1707.01348 [astro-ph.IM]} \BibitemShut {NoStop}%
\bibitem [{\citenamefont {Garcia-Bellido}\ \emph {et~al.}(2021)\citenamefont
  {Garcia-Bellido}, \citenamefont {Murayama},\ and\ \citenamefont
  {White}}]{Garcia-Bellido:2021zgu}%
  \BibitemOpen
  \bibfield  {author} {\bibinfo {author} {\bibfnamefont {J.}~\bibnamefont
  {Garcia-Bellido}}, \bibinfo {author} {\bibfnamefont {H.}~\bibnamefont
  {Murayama}}, \ and\ \bibinfo {author} {\bibfnamefont {G.}~\bibnamefont
  {White}},\ }\href@noop {} {\  (\bibinfo {year} {2021})},\ \Eprint
  {http://arxiv.org/abs/2104.04778} {arXiv:2104.04778 [hep-ph]} \BibitemShut
  {NoStop}%
\bibitem [{\citenamefont {Maggiore}\ \emph {et~al.}(2020)\citenamefont
  {Maggiore} \emph {et~al.}}]{Maggiore:2019uih}%
  \BibitemOpen
  \bibfield  {author} {\bibinfo {author} {\bibfnamefont {M.}~\bibnamefont
  {Maggiore}} \emph {et~al.},\ }\href {\doibase 10.1088/1475-7516/2020/03/050}
  {\bibfield  {journal} {\bibinfo  {journal} {JCAP}\ }\textbf {\bibinfo
  {volume} {03}},\ \bibinfo {pages} {050} (\bibinfo {year} {2020})},\ \Eprint
  {http://arxiv.org/abs/1912.02622} {arXiv:1912.02622 [astro-ph.CO]}
  \BibitemShut {NoStop}%
\bibitem [{\citenamefont {Inomata}\ and\ \citenamefont
  {Nakama}(2019)}]{Inomata:2018epa}%
  \BibitemOpen
  \bibfield  {author} {\bibinfo {author} {\bibfnamefont {K.}~\bibnamefont
  {Inomata}}\ and\ \bibinfo {author} {\bibfnamefont {T.}~\bibnamefont
  {Nakama}},\ }\href {\doibase 10.1103/PhysRevD.99.043511} {\bibfield
  {journal} {\bibinfo  {journal} {Phys. Rev. D}\ }\textbf {\bibinfo {volume}
  {99}},\ \bibinfo {pages} {043511} (\bibinfo {year} {2019})},\ \Eprint
  {http://arxiv.org/abs/1812.00674} {arXiv:1812.00674 [astro-ph.CO]}
  \BibitemShut {NoStop}%
\bibitem [{\citenamefont {Reitze}\ \emph {et~al.}(2019)\citenamefont {Reitze}
  \emph {et~al.}}]{Reitze:2019iox}%
  \BibitemOpen
  \bibfield  {author} {\bibinfo {author} {\bibfnamefont {D.}~\bibnamefont
  {Reitze}} \emph {et~al.},\ }\href@noop {} {\bibfield  {journal} {\bibinfo
  {journal} {Bull. Am. Astron. Soc.}\ }\textbf {\bibinfo {volume} {51}},\
  \bibinfo {pages} {035} (\bibinfo {year} {2019})},\ \Eprint
  {http://arxiv.org/abs/1907.04833} {arXiv:1907.04833 [astro-ph.IM]}
  \BibitemShut {NoStop}%
\bibitem [{\citenamefont {Janssen}\ \emph {et~al.}(2015)\citenamefont {Janssen}
  \emph {et~al.}}]{Janssen:2014dka}%
  \BibitemOpen
  \bibfield  {author} {\bibinfo {author} {\bibfnamefont {G.}~\bibnamefont
  {Janssen}} \emph {et~al.},\ }\href {\doibase 10.22323/1.215.0037} {\bibfield
  {journal} {\bibinfo  {journal} {PoS}\ }\textbf {\bibinfo {volume}
  {AASKA14}},\ \bibinfo {pages} {037} (\bibinfo {year} {2015})},\ \Eprint
  {http://arxiv.org/abs/1501.00127} {arXiv:1501.00127 [astro-ph.IM]}
  \BibitemShut {NoStop}%
\bibitem [{\citenamefont {Kasuya}\ and\ \citenamefont
  {Kawasaki}(2000{\natexlab{b}})}]{Kasuya:2000sc}%
  \BibitemOpen
  \bibfield  {author} {\bibinfo {author} {\bibfnamefont {S.}~\bibnamefont
  {Kasuya}}\ and\ \bibinfo {author} {\bibfnamefont {M.}~\bibnamefont
  {Kawasaki}},\ }\href {\doibase 10.1103/PhysRevLett.85.2677} {\bibfield
  {journal} {\bibinfo  {journal} {Phys. Rev. Lett.}\ }\textbf {\bibinfo
  {volume} {85}},\ \bibinfo {pages} {2677} (\bibinfo {year}
  {2000}{\natexlab{b}})},\ \Eprint {http://arxiv.org/abs/hep-ph/0006128}
  {arXiv:hep-ph/0006128} \BibitemShut {NoStop}%
\bibitem [{\citenamefont {Punturo}\ \emph {et~al.}(2010)\citenamefont {Punturo}
  \emph {et~al.}}]{Punturo:2010zz}%
  \BibitemOpen
  \bibfield  {author} {\bibinfo {author} {\bibfnamefont {M.}~\bibnamefont
  {Punturo}} \emph {et~al.},\ }\href {\doibase 10.1088/0264-9381/27/19/194002}
  {\bibfield  {journal} {\bibinfo  {journal} {Class. Quant. Grav.}\ }\textbf
  {\bibinfo {volume} {27}},\ \bibinfo {pages} {194002} (\bibinfo {year}
  {2010})}\BibitemShut {NoStop}%
\bibitem [{\citenamefont {Aggarwal}\ \emph {et~al.}(2020)\citenamefont
  {Aggarwal} \emph {et~al.}}]{Aggarwal:2020olq}%
  \BibitemOpen
  \bibfield  {author} {\bibinfo {author} {\bibfnamefont {N.}~\bibnamefont
  {Aggarwal}} \emph {et~al.},\ }\href@noop {} {\  (\bibinfo {year} {2020})},\
  \Eprint {http://arxiv.org/abs/2011.12414} {arXiv:2011.12414 [gr-qc]}
  \BibitemShut {NoStop}%
\bibitem [{\citenamefont {Khlopov}\ and\ \citenamefont
  {Linde}(1984)}]{Khlopov:1984pf}%
  \BibitemOpen
  \bibfield  {author} {\bibinfo {author} {\bibfnamefont {M.~Y.}\ \bibnamefont
  {Khlopov}}\ and\ \bibinfo {author} {\bibfnamefont {A.~D.}\ \bibnamefont
  {Linde}},\ }\href {\doibase 10.1016/0370-2693(84)91656-3} {\bibfield
  {journal} {\bibinfo  {journal} {Phys. Lett. B}\ }\textbf {\bibinfo {volume}
  {138}},\ \bibinfo {pages} {265} (\bibinfo {year} {1984})}\BibitemShut
  {NoStop}%
\bibitem [{\citenamefont {Pradler}\ and\ \citenamefont
  {Steffen}(2007)}]{Pradler:2006hh}%
  \BibitemOpen
  \bibfield  {author} {\bibinfo {author} {\bibfnamefont {J.}~\bibnamefont
  {Pradler}}\ and\ \bibinfo {author} {\bibfnamefont {F.~D.}\ \bibnamefont
  {Steffen}},\ }\href {\doibase 10.1016/j.physletb.2007.02.072} {\bibfield
  {journal} {\bibinfo  {journal} {Phys. Lett. B}\ }\textbf {\bibinfo {volume}
  {648}},\ \bibinfo {pages} {224} (\bibinfo {year} {2007})},\ \Eprint
  {http://arxiv.org/abs/hep-ph/0612291} {arXiv:hep-ph/0612291} \BibitemShut
  {NoStop}%
\bibitem [{\citenamefont {Kawasaki}\ \emph {et~al.}(2008)\citenamefont
  {Kawasaki}, \citenamefont {Kohri}, \citenamefont {Moroi},\ and\ \citenamefont
  {Yotsuyanagi}}]{Kawasaki:2008qe}%
  \BibitemOpen
  \bibfield  {author} {\bibinfo {author} {\bibfnamefont {M.}~\bibnamefont
  {Kawasaki}}, \bibinfo {author} {\bibfnamefont {K.}~\bibnamefont {Kohri}},
  \bibinfo {author} {\bibfnamefont {T.}~\bibnamefont {Moroi}}, \ and\ \bibinfo
  {author} {\bibfnamefont {A.}~\bibnamefont {Yotsuyanagi}},\ }\href {\doibase
  10.1103/PhysRevD.78.065011} {\bibfield  {journal} {\bibinfo  {journal} {Phys.
  Rev. D}\ }\textbf {\bibinfo {volume} {78}},\ \bibinfo {pages} {065011}
  (\bibinfo {year} {2008})},\ \Eprint {http://arxiv.org/abs/0804.3745}
  {arXiv:0804.3745 [hep-ph]} \BibitemShut {NoStop}%
\bibitem [{\citenamefont {Arya}\ \emph {et~al.}(2017)\citenamefont {Arya},
  \citenamefont {Mahajan},\ and\ \citenamefont {Rangarajan}}]{Arya:2016fnf}%
  \BibitemOpen
  \bibfield  {author} {\bibinfo {author} {\bibfnamefont {R.}~\bibnamefont
  {Arya}}, \bibinfo {author} {\bibfnamefont {N.}~\bibnamefont {Mahajan}}, \
  and\ \bibinfo {author} {\bibfnamefont {R.}~\bibnamefont {Rangarajan}},\
  }\href {\doibase 10.1016/j.physletb.2017.06.038} {\bibfield  {journal}
  {\bibinfo  {journal} {Phys. Lett. B}\ }\textbf {\bibinfo {volume} {772}},\
  \bibinfo {pages} {258} (\bibinfo {year} {2017})},\ \Eprint
  {http://arxiv.org/abs/1608.03386} {arXiv:1608.03386 [astro-ph.CO]}
  \BibitemShut {NoStop}%
\bibitem [{\citenamefont {Kawasaki}\ \emph {et~al.}(2018)\citenamefont
  {Kawasaki}, \citenamefont {Kohri}, \citenamefont {Moroi},\ and\ \citenamefont
  {Takaesu}}]{Kawasaki:2017bqm}%
  \BibitemOpen
  \bibfield  {author} {\bibinfo {author} {\bibfnamefont {M.}~\bibnamefont
  {Kawasaki}}, \bibinfo {author} {\bibfnamefont {K.}~\bibnamefont {Kohri}},
  \bibinfo {author} {\bibfnamefont {T.}~\bibnamefont {Moroi}}, \ and\ \bibinfo
  {author} {\bibfnamefont {Y.}~\bibnamefont {Takaesu}},\ }\href {\doibase
  10.1103/PhysRevD.97.023502} {\bibfield  {journal} {\bibinfo  {journal} {Phys.
  Rev. D}\ }\textbf {\bibinfo {volume} {97}},\ \bibinfo {pages} {023502}
  (\bibinfo {year} {2018})},\ \Eprint {http://arxiv.org/abs/1709.01211}
  {arXiv:1709.01211 [hep-ph]} \BibitemShut {NoStop}%
\bibitem [{\citenamefont {Eberl}\ \emph {et~al.}(2021)\citenamefont {Eberl},
  \citenamefont {Gialamas},\ and\ \citenamefont {Spanos}}]{Eberl:2020fml}%
  \BibitemOpen
  \bibfield  {author} {\bibinfo {author} {\bibfnamefont {H.}~\bibnamefont
  {Eberl}}, \bibinfo {author} {\bibfnamefont {I.~D.}\ \bibnamefont {Gialamas}},
  \ and\ \bibinfo {author} {\bibfnamefont {V.~C.}\ \bibnamefont {Spanos}},\
  }\href {\doibase 10.1103/PhysRevD.103.075025} {\bibfield  {journal} {\bibinfo
   {journal} {Phys. Rev. D}\ }\textbf {\bibinfo {volume} {103}},\ \bibinfo
  {pages} {075025} (\bibinfo {year} {2021})},\ \Eprint
  {http://arxiv.org/abs/2010.14621} {arXiv:2010.14621 [hep-ph]} \BibitemShut
  {NoStop}%
\bibitem [{\citenamefont {Blinov}\ \emph {et~al.}(2021)\citenamefont {Blinov},
  \citenamefont {Krnjaic},\ and\ \citenamefont {Li}}]{Blinov:2021mdk}%
  \BibitemOpen
  \bibfield  {author} {\bibinfo {author} {\bibfnamefont {N.}~\bibnamefont
  {Blinov}}, \bibinfo {author} {\bibfnamefont {G.}~\bibnamefont {Krnjaic}}, \
  and\ \bibinfo {author} {\bibfnamefont {S.~W.}\ \bibnamefont {Li}},\
  }\href@noop {} {\  (\bibinfo {year} {2021})},\ \Eprint
  {http://arxiv.org/abs/2108.11386} {arXiv:2108.11386 [hep-ph]} \BibitemShut
  {NoStop}%
\bibitem [{\citenamefont {Barrie}\ \emph {et~al.}(2021)\citenamefont {Barrie},
  \citenamefont {Han},\ and\ \citenamefont {Murayama}}]{Barrie:2021mwi}%
  \BibitemOpen
  \bibfield  {author} {\bibinfo {author} {\bibfnamefont {N.~D.}\ \bibnamefont
  {Barrie}}, \bibinfo {author} {\bibfnamefont {C.}~\bibnamefont {Han}}, \ and\
  \bibinfo {author} {\bibfnamefont {H.}~\bibnamefont {Murayama}},\ }\href@noop
  {} {\  (\bibinfo {year} {2021})},\ \Eprint {http://arxiv.org/abs/2106.03381}
  {arXiv:2106.03381 [hep-ph]} \BibitemShut {NoStop}%
\bibitem [{\citenamefont {Kawasaki}\ and\ \citenamefont
  {Ueda}(2021)}]{Kawasaki:2020xyf}%
  \BibitemOpen
  \bibfield  {author} {\bibinfo {author} {\bibfnamefont {M.}~\bibnamefont
  {Kawasaki}}\ and\ \bibinfo {author} {\bibfnamefont {S.}~\bibnamefont
  {Ueda}},\ }\href {\doibase 10.1088/1475-7516/2021/04/049} {\bibfield
  {journal} {\bibinfo  {journal} {JCAP}\ }\textbf {\bibinfo {volume} {04}},\
  \bibinfo {pages} {049} (\bibinfo {year} {2021})},\ \Eprint
  {http://arxiv.org/abs/2011.10397} {arXiv:2011.10397 [hep-ph]} \BibitemShut
  {NoStop}%
\bibitem [{\citenamefont {Cline}\ \emph {et~al.}(2020)\citenamefont {Cline},
  \citenamefont {Puel},\ and\ \citenamefont {Toma}}]{Cline:2019fxx}%
  \BibitemOpen
  \bibfield  {author} {\bibinfo {author} {\bibfnamefont {J.~M.}\ \bibnamefont
  {Cline}}, \bibinfo {author} {\bibfnamefont {M.}~\bibnamefont {Puel}}, \ and\
  \bibinfo {author} {\bibfnamefont {T.}~\bibnamefont {Toma}},\ }\href {\doibase
  10.1103/PhysRevD.101.043014} {\bibfield  {journal} {\bibinfo  {journal}
  {Phys. Rev. D}\ }\textbf {\bibinfo {volume} {101}},\ \bibinfo {pages}
  {043014} (\bibinfo {year} {2020})},\ \Eprint
  {http://arxiv.org/abs/1909.12300} {arXiv:1909.12300 [hep-ph]} \BibitemShut
  {NoStop}%
\bibitem [{\citenamefont {Yamada}(2016)}]{Yamada:2015xyr}%
  \BibitemOpen
  \bibfield  {author} {\bibinfo {author} {\bibfnamefont {M.}~\bibnamefont
  {Yamada}},\ }\href {\doibase 10.1103/PhysRevD.93.083516} {\bibfield
  {journal} {\bibinfo  {journal} {Phys. Rev. D}\ }\textbf {\bibinfo {volume}
  {93}},\ \bibinfo {pages} {083516} (\bibinfo {year} {2016})},\ \Eprint
  {http://arxiv.org/abs/1511.05974} {arXiv:1511.05974 [hep-ph]} \BibitemShut
  {NoStop}%
\bibitem [{\citenamefont {Harigaya}\ \emph {et~al.}(2014)\citenamefont
  {Harigaya}, \citenamefont {Kamada}, \citenamefont {Kawasaki}, \citenamefont
  {Mukaida},\ and\ \citenamefont {Yamada}}]{Harigaya:2014tla}%
  \BibitemOpen
  \bibfield  {author} {\bibinfo {author} {\bibfnamefont {K.}~\bibnamefont
  {Harigaya}}, \bibinfo {author} {\bibfnamefont {A.}~\bibnamefont {Kamada}},
  \bibinfo {author} {\bibfnamefont {M.}~\bibnamefont {Kawasaki}}, \bibinfo
  {author} {\bibfnamefont {K.}~\bibnamefont {Mukaida}}, \ and\ \bibinfo
  {author} {\bibfnamefont {M.}~\bibnamefont {Yamada}},\ }\href {\doibase
  10.1103/PhysRevD.90.043510} {\bibfield  {journal} {\bibinfo  {journal} {Phys.
  Rev. D}\ }\textbf {\bibinfo {volume} {90}},\ \bibinfo {pages} {043510}
  (\bibinfo {year} {2014})},\ \Eprint {http://arxiv.org/abs/1404.3138}
  {arXiv:1404.3138 [hep-ph]} \BibitemShut {NoStop}%
\bibitem [{\citenamefont {Garcia}\ and\ \citenamefont
  {Olive}(2013)}]{Garcia:2013bha}%
  \BibitemOpen
  \bibfield  {author} {\bibinfo {author} {\bibfnamefont {M.~A.~G.}\
  \bibnamefont {Garcia}}\ and\ \bibinfo {author} {\bibfnamefont {K.~A.}\
  \bibnamefont {Olive}},\ }\href {\doibase 10.1088/1475-7516/2013/09/007}
  {\bibfield  {journal} {\bibinfo  {journal} {JCAP}\ }\textbf {\bibinfo
  {volume} {09}},\ \bibinfo {pages} {007} (\bibinfo {year} {2013})},\ \Eprint
  {http://arxiv.org/abs/1306.6119} {arXiv:1306.6119 [hep-ph]} \BibitemShut
  {NoStop}%
\bibitem [{\citenamefont {Kusenko}(1999)}]{Kusenko:1998yi}%
  \BibitemOpen
  \bibfield  {author} {\bibinfo {author} {\bibfnamefont {A.}~\bibnamefont
  {Kusenko}},\ }\href {\doibase 10.1063/1.59410} {\bibfield  {journal}
  {\bibinfo  {journal} {AIP Conf. Proc.}\ }\textbf {\bibinfo {volume} {478}},\
  \bibinfo {pages} {312} (\bibinfo {year} {1999})},\ \Eprint
  {http://arxiv.org/abs/hep-ph/9901353} {arXiv:hep-ph/9901353} \BibitemShut
  {NoStop}%
\end{thebibliography}%
\appendix 
\subsection{Explicit Potentials}\label{ap:potentials}

To keep the discussion as general as possible, we have avoided specifying a potential.  In this appendix, we discuss Q-balls in both gauge-mediated and gravity-mediated scenarios.

In the gauge-mediated supersymmetric scenario the potential is
\begin{equation}
    V(\Phi ) = m^4 \log \left( 1+ \frac{|\Phi|^2}{m^2} \right) + \frac{1}{\Lambda ^2} |\Phi|^6,
    \label{eq:potential}
\end{equation}
plus a small CP-violating term.  When the second term is negligible, the resulting Q-balls are thick wall Q-ball~\cite{Multamaki:1999an}, for which the analysis presented here is inapplicable.  However, scaling arguments favor a Q-ball domination epoch.  Fragmentation tends to produce one large Q-ball in each Hubble volume~\cite{Kasuya:1999wu}, resulting in a sharply peaked mass distribution at large masses, which tend to be long-lived.  Furthermore, the radius now scales as $Q^{1 \slash 4}$, and after accounting to the scalings of the VEV and energy per unit charge with $Q$, we expect the Q-ball decay rate to scale as $Q^{1 \slash 4}$.  This is suppressed compared to the thin-wall rate, and therefore, the Q-balls will tend to be longer-lived.  $\Gamma_{\rm Q-ball} \slash HQ$ then scales as $Q^{-3 \slash 4}$, which increases as the charge decreases, leading to the rapid matter-to-radiation transition.  We plan to address this scenario more fully in future work.

When the second term in \eqref{eq:potential} is not negligible, then the resulting Q-balls are in the thin wall regime, although since the energy per charge $\omega$ is independent of the charge $Q$, the equilibrium state is not one Q-ball per Hubble volume.

Alternatively, one can consider gravity-mediated SUSY breaking, in which case the Affleck-Dine condensate has the potential 
\begin{equation}
V(\Phi) = m^2 |\Phi|^2 \left( 1 + K \log \left( \dfrac{|\Phi|^2}{m^2} \right)\right) + \dfrac{1}{\Lambda^2} |\Phi|^6,
\label{eq:potential2}
\end{equation}
where $\Lambda$ is an effective scale for the higher-dimensional operator, $m$ is the mass of the scalar field, and $K \approx -0.01$ to $-0.1$ is a one-loop correction.  Since $K<0$, Q-balls can be formed, and in the thin wall limit, the VEV inside the Q-ball is~\cite{Coleman:1985ki}
\begin{equation}
v \approx \left( \Lambda M_{\rm Pl} \sqrt{ \dfrac{|K|}{2}} \right)^{1 \slash 2},
\end{equation}
from which an expression for $\omega$ can be found.

In Table \ref{tab:my_label3} we show one choice of parameters for the gauge-mediated potential (left) and gravity-mediated potential (right) for each benchmark set of parameters discussed in the text.  That is, in each row the Q-balls produced have the same VEV $v$ and energy-per-unit-charge $\omega$ as in corresponding row in Table \ref{tab:my_label}.  For the gauge-mediated scenario, we have ensured that the sixth order term is relevant so that we are in the thin-wall regime.  We see that in all cases the scale of the effective operator, $\Lambda$, is well above the reheating temperature.

\begin{table}[t]
    \centering
    \begin{tabular}{|c|c|c|c|} \hline
\multicolumn{2}{|c|}{Gauge-Mediated} & 
\multicolumn{2}{c|}{Gravity-Mediated} \\ \hline
$ m $ (GeV) & $\Lambda$ (GeV)&$ m $ (GeV) & $\Lambda$ (GeV) \\ \hline
$6.07 \times 10^7$ & $4.54 \times 10^{15}$ & $2.16 \times 10^6$ & $4.23 \times 10^{15} $\\
$1.76 \times 10^7$ & $1.10 \times 10^{13}$ & $1.13 \times 10^6$ & $2.07 \times 10^{13}$ \\
$3.61 \times 10^7$ & $1.35 \times 10^{14}$  &$1.63 \times 10^6$ & $2.01 \times 10^{14} $ \\ \hline 
    \end{tabular}
    \caption{Parameters for the potentials \eqref{eq:potential} and \eqref{eq:potential2} which produce Q-balls corresponding to our three benchmark points.  The first row corresponds to our first benchmark in Table \ref{tab:my_label}, the second row corresponds the second benchmark, and the third corresponds to the third.  For the gravity-mediated scenario, we have fixed $k=-0.05$.}
    \label{tab:my_label3}
\end{table}

\subsection{Gravitational wave spectrum}\label{ap:GW}

We here outline the gravitational wave spectrum from an instantaneous transition from matter to radiation domination, following Ref~\cite{Inomata:2019ivs}.  We define the following functions:
\begin{align}
& s_0 (x,x_{\rm max})= \left( \Theta \left[ \frac{2 x_{\rm max}}{1+\sqrt{3}}-x \right]   \right. + \left. \left(2 \frac{x_{\rm max}}{x} - \sqrt{3} \right) \right. \nonumber \\ & \left. \Theta \left[ x- 2 \frac{x_{\rm max}}{1+ \sqrt{3}} \right] \Theta \left[ 2 \frac{x_{\rm max}}{\sqrt{3}} -x \right] \right)  \times  \Theta \left[ 2 \frac{x_{\rm max}}{ \sqrt{3}} - x \right] ,
\end{align}
along with 
\begin{align}
    S_i (x) &= \int _0 ^x dz \frac{\sin z}{z} , \qquad C_i (x) = - \int _x ^\infty dz \frac{\cos z}{z} \ .  
\end{align}
The gravitational wave spectrum involves a resonant contribution, $\Omega _{\rm res}$, an infrared contribution, $\Omega _{\rm IR}$, and a non-resonant UV contribution, $\Omega _{\rm UV}$, such that the total spectrum is given by
\begin{align}
&    \Omega _{\rm GW}(x,x_{\rm max}) = \Omega _{\rm res} + \nonumber \\ 
    &3 A_s^2 x_{\rm max}^8 \frac{4 C_i[\frac{x}{2}]^2+(\pi - 2 S_i [\frac{x}{2}])^2}{2^{17+2 n_s}625(3+2n_s)} \left( \frac{2 x_{\rm max}}{x} -1 \right) ^{2n_s} \nonumber \\ & \times \left( \frac{x}{x_\ast} \right) ^{2 (n_s -1)} \left( \Omega _{\rm IR} \Theta [x_{\rm max}-x] + \Omega _{\rm UV} \Theta \left[ x-x_{\rm max} \right] \right), \nonumber \\
\end{align}
where we define $X \equiv x \slash x_{\rm max}$ to write 
\begin{align}
&    \Omega _{\rm IR}= \frac{1}{(2+n_s)(3+n_s)(4+n_s)(5+2n_s)(7+2n_s)} \nonumber \\ 
&\times  \left( 1536-6144 X + (7168-1920n_s-256 n_s^2) X^2 \right. \nonumber \\ 
& \left. + (5760 n_s +768 n_s^2)X^3 \right. \nonumber \\ & \left. +(1328 n_s +3056 n_s^2 + 832 n_s^3 +64 n_s^4) X^4 \right. \nonumber \\ 
& - \left. (7168 +12256 n_s +7392 n_s^2 +1664 n_s^3 +128 n_s^4)X^5 \right. \nonumber \\ 
& + \left. (7392 + 10992 n_s +5784 n_s^2 + 1248 n_s^3 +96 n_s^4) X^6 \right. \nonumber \\ 
& - \left. (2784+3904 n_s +1960 n_s^2 +416 n_s^3 + 32 n_s^4) X^7 \right. \nonumber \\ 
& + \left. (370 + 503 n_s +247 n_s^2 + 52 n_s^3 + 4 n_s^4) X^8 \right. \nonumber \\ 
& - \left. 256 \left(1- X\right)^6 \left[(6+ 6(2+n_s) X \right. \right. \nonumber \\ 
& \left. \left. +(2+n_s)(5+2n_s)X^2\right]\left(1-\frac{X}{2- X}\right)^{2 n_s} \right),
\end{align}
and
\begin{align}
&   \Omega _{\rm UV} = 2 \left( 2 - X\right)^4 \Gamma [4+2n_s]\nonumber \\ 
    & \times \left( \dfrac{X^4}{\Gamma [5+2n_s]} - \frac{4 X^2 \left(2-X \right)^2}{\Gamma [7+2n_s]}  + \frac{24 \left( 2-X\right)^4}{\Gamma [9+2n_s]} \right),
\end{align}
\begin{align}
    \Omega _{\rm res} &= \frac{2.3 \sqrt{3} 3^{n_s}}{625 \times 2^{13+2 n_s}} x^7 X^{2(n_s-1)} A_s^2 s_0(x,x_{\rm max}) \nonumber \\ & \times \left( 4 _2F_1 [\frac{1}{2} , 1-n_s , \frac{3}{2},\frac{s_0(x,x_{\rm max})^2}{3}] \right. \nonumber \\ &-  \left.  3 _2F_1 [\frac{1}{2} , -n_s , \frac{3}{2},\frac{s_0(x,x_{\rm max})^2}{3}] \right. \nonumber \\ & \left. - s_0(x,x_{\rm max})^2  {}_2F_1 [\frac{3}{2} , -n_s , \frac{5}{2},\frac{s_0(x,x_{\rm max})^2}{3}]  \right) \nonumber \\ 
\end{align}
where $ {}_2 F_1$ is the hypergeometric function.
To convert to a frequency spectrum, simply take $x_\ast = 1/k_\ast \eta _r$ and \begin{align}  &\Omega _{\rm GW} (f) h^2  = 0.39 h^2 \Omega _r \nonumber \\ &\quad \times  \Omega _{\rm GW}  \left[4.1 \times 10^{-24} \left(\frac{\eta _r}{\rm GeV^{-1}}\right) f,\frac{T_{\rm eq}}{T_{\rm dec}} \right].  \end{align}
\end{document}